\documentclass[prl,aps,twocolumn,preprintnumbers,showpacs,showkeys,superscriptaddress]{revtex4}
\usepackage{epsfig,amssymb,amsfonts,verbatim}

\def\bfl{\begin{flushleft}}
\def\efl{\end{flushleft}}
\def\bfr{\begin{flushright}}
\def\efr{\end{flushright}}
\def\bc{\begin{center}}
\def\ec{\end{center}}
\def\be{\begin{equation}}
\def\ee{\end{equation}}
\def\ba{\begin{eqnarray}}
\def\ea{\end{eqnarray}}
\def\baa#1{\begin{array}{#1}}
\def\eaa{\end{array}}
\def\bw{\begin{widetext}}
\def\ew{\end{widetext}}
\def\nn{\nonumber }
\def\lb#1{\label{#1}}

\begin{document}

\title{Ionization energy based Fermi-Dirac statistics}

\author{Andrew Das Arulsamy}
\affiliation{Condensed Matter Group, Division of Exotic Matter,
No. 22, Jalan Melur 14, Taman Melur, 68000 Ampang, Selangor DE,
Malaysia}

%\date{1$^{st}$ January 2003}
%\date{~Received: 12 Jan 2001 [LANL] ~}
%\date{~Received: 26 May 2000 [PRL], 1 June 2000 [LANL] ~}
\date{\today}
%\small
%\scriptsize%\footnotesize

\begin{abstract}
Quantitative differences of Lagrange multipliers between standard
Fermi-Dirac statistics (FDS) and Ionization energy ($E_I$) based
FDS (iFDS) are analyzed in detail. It is shown here that iFDS is
degenerate and its total energy remains the same with the standard
FDS. The total energy can be obtained by recasting the $E_I$ in
quantized form, as required by the new restrictive condition.
Basically, iFDS provides an alternative route to calculate the
Fermions' distribution spectrum.
\end{abstract}

 \maketitle

%\narrowtext

\subsection{1. Introduction}\lb{s-in}

iFDS has been used recently to understand the evolution of
resistivity curves with doping and temperature ($T$) in a wide
variety of electronic
matter~\cite{andrew1,andrew2,andrew3,andrew4,andrew5,andrew6,arulsamy1,arulsamy2}.
Such understanding and iFDS's prediction can only be captured by
introducing the parameter, $E_I$. Here, the Lagrange multipliers,
degeneracy and the total energy requirement associated with $E_I$
in iFDS is discussed in detail. Both FDS and iFDS are for the
half-integral spin particles such as electrons and holes. Its
total wave function, $\Psi$ has to be antisymmetric in order to
satisfy quantum-mechanical symmetry requirement. Under such
condition, interchange of any 2 particles ($A$ and $B$) of
different states, $\psi_i$ and $\psi_j$ ($j$ $\ne$ $i$) will
result in change of sign, hence the wave function for Fermions is
in the form of

\begin {eqnarray}
\Psi_{i,j}(C_A,C_B) = \psi_i(C_A)\psi_j(C_B) -
\psi_i(C_B)\psi_j(C_A). \label{eq:1}
\end {eqnarray}

The negative sign in Eq.~(\ref{eq:1}) that fulfils antisymmetric
requirement is actually due to one of the eigenvalue of exchange
operator~\cite{griffiths6}, {\bf P} = $-$1. The other eigenvalue,
{\bf P} = $+$1 is for Bosons. $C_A$ and $C_B$ denote all the
necessary cartesian coordinates of the particles $A$ and $B$
respectively. Equation~(\ref{eq:1}) is nothing but Pauli's
exclusion principle. The one-particle energies $E_1$, $E_2$, ...,
$E_m$ for the corresponding one-particle quantum states $q_1$,
$q_2$, ..., $q_m$ can be rewritten as ($E_{is}$ $\pm$ $E_I)_1$,
($E_{is}$ $\pm$ $E_I)_2$, ..., ($E_{is}$ $\pm$ $E_I)_m$. Note here
that $E_{is}$ = $E_{initial~state}$. It is also important to
realize that $E_{is}$ + $E_I$ = $E_{electrons}$ and $E_{is}$ $-$
$E_I$ = $E_{holes}$. Subsequently, the latter ($E_{is}$ $\pm$
$E_I)_i$ version where $i$ = 1, 2, ..., $m$ with $E_I$ as an
additional inclusion will be used to derive iFDS and its Lagrange
multipliers. This $\pm E_I$ is inserted carefully to justify that
an electron to occupy a higher state $N$ from initial state $M$ is
more probable than from initial state $L$ if condition $E_I(M)$
$<$ $E_I(L)$ at certain $T$ is satisfied. As for a hole to occupy
a lower state $M$ from initial state $N$ is more probable than to
occupy state $L$ if the same condition above is satisfied.
$E_{is}$ is the energy of a particle in a given system at a
certain initial state and ranges from $+\infty$ to 0 for electrons
and 0 to $-\infty$ for holes. In contrast, standard FDS only
requires $E_i$ ($i$ = 1, 2, ..., $m$) as the energy of a particle
at a certain state.

\subsection{2. Theoretical details}\lb{s-in}

Denoting $n$ as the total number of particles with $n_1$ particles
with energy ($E_{is}$ $\pm$ $E_I)_1$, $n_2$ particles with energy
($E_{is}$ $\pm$ $E_I)_2$ and so on implies that $n$ = $n_1$ +
$n_2$ + ... + $n_m$. As a consequence, the number of ways for
$q_1$ quantum states to be arranged among $n_1$ particles is given
as

\begin {eqnarray}
P(n_1,q_1) = \frac{q_1!}{n_1!(q_1 - n_1)!}.\label{eq:2}
\end {eqnarray}

Now it is easy to enumerate the total number of ways for $q$
quantum states ($q$ = $q_1$ + $q_2$ + ... + $q_m$) to be arranged
among $n$ particles, which is

\begin {eqnarray}
P(n,q) = \prod\limits_{i=1}^{\infty} \frac{q_i!}{n_i!(q_i - n_i)!} .\label{eq:3}
\end {eqnarray}

The most probable configuration at certain $T$ can be obtained by
maximizing $P(n,q)$ subject to the restrictive conditions

\begin {eqnarray}
&&\sum_i^{\infty} n_i = n, \sum_i^{\infty} dn_i = 0. \label{eq:4}
\end {eqnarray}

\begin {eqnarray}
&&\sum_i^{\infty} (E_{is}\pm E_I)_i n_i = E, \sum_i^{\infty}
(E_{is}\pm E_I)_i dn_i = 0.\label{eq:5}
\end {eqnarray}

The method of Lagrange multipliers~\cite{griffiths6} can be
employed to maximize Eq.~(\ref{eq:3}). Hence, a new function,
$F(x_1, x_2, ...\mu, \lambda,...)$ = $f + \mu f_1 + \lambda f_2$
+... is introduced and all its derivatives are set to zero

\begin {eqnarray}
\frac{\partial F}{\partial x_n} = 0;~~~ \frac{\partial F}{\partial
\mu} = 0;~~~ \frac{\partial F}{\partial \lambda} = 0.\label{eq:6}
\end {eqnarray}

As such, one can let the new function in the form of

\begin {eqnarray}
F = \ln P + \mu \sum_i^{\infty} dn_i + \lambda \sum_i^{\infty}
(E_{is}\pm E_I)_i dn_i.\label{eq:7}
\end {eqnarray}

After applying Stirling's approximation, $\partial F$/$\partial n_i$ can be written as

\begin {eqnarray}
\frac {\partial F}{\partial n_i}&& = \ln (q_i - n_i) - \ln n_i +
\mu + \lambda (E_{is}\pm E_I)_i \nn \\&& = 0.\label{eq:8}
\end {eqnarray}

Thus, the Fermi-Dirac statistics based on ionization energy is simply given by

\begin {eqnarray}
\frac {n_i}{q_i} = \frac{1}{\exp [\mu + \lambda (E_{is}\pm E_I)_i]
+ 1}.\label{eq:9}
\end {eqnarray}

%\subsection{2.2. Lagrange multipliers}\lb{s-eqs}

Importantly, the total energy, $E$ in iFDS can be obtained from
Eq.~(\ref{eq:5}), which is

\begin {eqnarray}
&E& = \sum_i^{\infty} (E_{is}\pm E_I)_i n_i \nn \\&& =
\sum_i^{\infty} \frac{\hbar^2}{2m}\big[{\bf k}^2_{is}\pm {\bf
k}^2_I\big]_i n_i \nn
\\&& = \frac{\hbar^2}{2m}\big[{\bf k}^2_{is}\pm {\bf
k}^2_I\big] = \frac{\hbar^2}{2m}{\bf k}^2.\label{eq:10}
\end {eqnarray}

$\textbf{k}_I$ = $\textbf{k}_{ionized~state}$, and the $\pm$ sign
is solely to indicate that the energy corresponds to electrons is
0 $\to$ $+\infty$ while 0 $\to$ $-\infty$ is for the holes, which
satisfy the particle-hole symmetry. Consequently,
Eq.~(\ref{eq:10}) also implies that iFDS does not violate the
degeneracy requirements. By utilizing Eq.~(\ref{eq:9}) and taking
$\exp[\mu + \lambda(E \pm E_I)]$ $\gg$ 1, one can arrive at the
probability function for electrons in an explicit form as

\begin{eqnarray}
f_e({\bf k}_{is}) = \exp \left[-\mu-\lambda\left(\frac{\hbar^2{\bf
k}_{is}^2}{2m}+E_I\right) \right], \label{eq:11}
\end{eqnarray}

Similarly, the probability function for the holes is given by

\begin{eqnarray}
f_h({\bf k}_{is}) = \exp\left[\mu + \lambda\left(\frac{\hbar^2{\bf
k}_{is}^2}{2m}-E_I\right) \right]. \label{eq:12}
\end{eqnarray}

The parameters $\mu$ and $\lambda$ are the Lagrange multipliers.
$\hbar$ $=$ $h/2\pi$, $h$ $=$ Planck constant and $m$ is the
charge carriers' mass. Note that $E$ has been substituted with
$\hbar^2{\bf k}^2/2m$. In the standard FDS, Eqs.~(\ref{eq:11})
and~(\ref{eq:12}) are simply given by, $f_e({\bf k})$ $=$
$\exp[-\mu-\lambda(\hbar^2{\bf k}^2/2m)]$ and $f_h({\bf k})$ $=$
$\exp[\mu+\lambda(\hbar^2{\bf k}^2/2m)]$. Equation~(\ref{eq:4})
can be rewritten by employing the 3D density of states' (DOS)
derivative, $dn$ $=$ $V{\bf k}_{is}^2d{\bf k}_{is}/2\pi^2$,
Eqs.~(\ref{eq:11}) and~(\ref{eq:12}), that eventually give

\begin{eqnarray}
&n& = \frac {V}{2\pi^2}e^{-\mu} \int\limits_0^\infty {\bf k}^2
\exp\bigg[-\lambda \frac{\hbar^2{\bf k}^2}{2m}\bigg] d{\bf k} \nn
\\&& = \frac {V}{2\pi^2}e^{-\mu} \int\limits_0^\infty {\bf
k}_{is}^2 \exp\bigg[-\lambda \frac{\hbar^2{\bf k}_{is}^2}{2m}
-\lambda \frac{\hbar^2{\bf k}_I^2}{2m} \bigg] d{\bf k}_{is} \nn
\\&& = \frac {V}{2\pi^2}e^{-\mu-\lambda E_I} \int\limits_0^\infty
{\bf k}_{is}^2 \exp\bigg[-\lambda \frac{\hbar^2{\bf
k}_{is}^2}{2m}\bigg] d{\bf k}_{is}, \label{eq:13}
\end{eqnarray}

\begin{eqnarray}
p & = & \frac {V}{2\pi^2}e^{\mu - \lambda E_I}
\int\limits_{-\infty}^0 {\bf k}_{is}^2 \exp\bigg[\lambda \frac
{\hbar^2{\bf k}_{is}^2}{2m}\bigg] d{\bf k}_{is}.\label{eq:14}
\end{eqnarray}

The respective solutions for Eqs.~(\ref{eq:13}) and~(\ref{eq:14})
are

\begin{eqnarray}
\mu + \lambda E_I & = &
-\ln\left[\frac{n}{V}\left(\frac{2\pi\lambda\hbar^2}{m}\right)^{3/2}
\right],\label{eq:15}
\end{eqnarray}

\begin{eqnarray}
\mu -\lambda E_I & = &
\ln\left[\frac{p}{V}\left(\frac{2\pi\lambda\hbar^2}{m}\right)^{3/2}
\right].\label{eq:16}
\end{eqnarray}

Note that Eqs.~(\ref{eq:15}) and~(\ref{eq:16}) simply imply that
$\mu_e(iFDS)$ $=$ $\mu(T=0)$ + $\lambda E_I$ and $\mu_h(iFDS)$ $=$
$\mu(T=0)$ $-$ $\lambda E_I$. In fact, $\mu(FDS)$ need to be
varied accordingly with doping, on the other hand, iFDS captures
the same variation due to doping with $\lambda E_I$ in which,
$\mu(T=0)$ is fixed to be a constant (independent of $T$ and
doping). Furthermore, using Eq.~(\ref{eq:5}), one can obtain

\begin{eqnarray}
E && = \frac {V\hbar^2}{4m\pi^2} e^{-\mu(FDS)}
\int\limits_0^\infty {\bf k}^4 \exp\bigg[-\lambda
\frac{\hbar^2{\bf k}^2}{2m}\bigg] d{\bf k} \nn
\\&& = \frac {V\hbar^2}{4m\pi^2} e^{-\mu(T=0)} \int\limits_0^\infty {\bf
k}_{is}^4 \exp\bigg[-\lambda \frac{\hbar^2{\bf k}_{is}^2}{2m}
-\lambda \frac{\hbar^2{\bf k}_I^2}{2m} \bigg] d{\bf k}_{is} \nn
\\&& = \frac {V\hbar^2}{4m\pi^2} e^{-\mu(T=0) -\lambda E_I}
\int\limits_0^\infty {\bf k}_{is}^4
\exp\bigg[-\lambda\frac{\hbar^2{\bf k}_{is}^2}{2m}\bigg]d{\bf
k}_{is} \nn
\\&& = \frac{3V}{2\lambda}e^{-\mu(T=0) -\lambda E_I}
\bigg[\frac{m}{2\pi\lambda\hbar^2}\bigg]^{3/2} \label{eq:17}
\\&& = \frac{3V}{2\lambda}e^{-\mu(FDS)}
\bigg[\frac{m}{2\pi\lambda\hbar^2}\bigg]^{3/2}. \label{eq:18}
\end{eqnarray}

Again, Eq.~(\ref{eq:17}) being equal to Eq.~(\ref{eq:18}) enable
one to surmise that the total energy considered in FDS and iFDS is
exactly the same. Quantitative comparison between
Eq.~(\ref{eq:17}) and with the energy of a 3D ideal gas, $E$ $=$
$3nk_BT/2$, after substituting Eq.~(\ref{eq:15}) into
Eq.~(\ref{eq:17}) will enable one to determine $\lambda$. It is
found that $\lambda$ remains the same as 1/$k_BT$.

\subsection{3. Discussion}\lb{s-in}

Recall that the $E_I$ here corresponds to the energy needed to
ionize an atom or ion in such a way that the electrons are excited
to an energy level distanced at $r$, not $\infty$. However, the
proportionality, $E_I(r=r)$ $\propto$ $E_I(r=\infty)$ is valid,
which has been used to describe the experimental data of strongly
correlated matter with minuscule substitutional
doping~\cite{andrew1,andrew2,andrew3,andrew4,andrew5,andrew6}.
Basically, at constant temperature ($T$ $>$ 0), FDS predicts the
distribution spectrum if $E$ is varied, relying on external inputs
such as band gap ($E_g$) and/or Fermi level ($E_F(T)$). On the
other hand, iFDS needs only $E_I$ as an external input to predict
the variation of $E$, without relying on $E_g$ and/or $E_F(T)$ at
all, and subsequently its distribution spectrum can be obtained as
well. Notice that $E_F$ comes into iFDS as $E_F^0$ = constant,
independent of $T$ and doping. $E_F^0$ denotes the Fermi level at
0 K. The $E_I$ is microscopically defined as~\cite{andrew6}

\begin {eqnarray}
\epsilon(0,\textbf{k}) = 1 +
\frac{\mathcal{K}_s^2}{\textbf{k}^2}\exp\big[\lambda^*(E_F^0-E_I)\big].\label{eq:19}
\end {eqnarray}

$\epsilon(0,\textbf{k})$ is the static dielectric function, while
$\mathcal{K}_s$ represents the Thomas-Fermi screening parameter.
$\lambda^* = (12\pi\epsilon_0/e^2)\verb"n"^2r_B$, $\epsilon_0$ and
\verb"n" are the permittivity of space and principal quantum
number respectively while $r_B$ denotes the Bohr radius. In fact,
iFDS and FDS take different approach in term of energy levels and
Fermions' excitations to arrive at the same distribution spectrum.
In simple words, iFDS is new in a sense that it gives one an
alternative route to obtain the Fermions' distribution spectrum in
which, FDS needs $E_F(T)$ and/or $E_g$ while iFDS needs only $E_I$
to arrive at the same distribution spectrum. Hence, based on the
accuracy of these input parameters, one can choose either FDS or
iFDS to be used for one's theoretical models.

\subsection{4. Conclusions}\lb{s-in}

In conclusion, the relationship between FDS and iFDS in term of
Lagrange multipliers has been derived and shown clearly. The total
energy considered in ionization energy based Fermi-Dirac
statistics is as same as the FDS. Actually, the total energy has
been recast into a fundamental form that consists of initial state
and ionized state energies. iFDS's prediction are also remarkable
in non-free-electron metals, namely High-$T_c$
superconductors~\cite{andrew1,andrew2,andrew3,andrew4,arulsamy1,arulsamy2},
feromagnets~\cite{andrew5} and ferroelectrics~\cite{andrew6}. As
such, one has the option whether to adopt FDS or iFDS based on
reliable input parameters in which, $E_F(T)$ and/or $E_g$
corresponds to FDS while $E_I$ is connected with iFDS.

\subsection*{Acknowledgments}\lb{s-in}

The author is grateful to Arulsamy Innasimuthu, Sebastiammal
Innasimuthu, Arokia Das Anthony and Cecily Arokiam of CMG-A for
their hospitality.

\end{document}